\documentstyle[preprint,aps]{revtex}
\begin{document}
\draft
\title{Experimental Evidence of a Haldane Gap in an $\bbox{S = 2}$ Quasi-linear Chain
 Antiferromagnet}
\author{G.E. Granroth and M.W. Meisel} 
\address{Department of Physics and Center for Ultralow Temperature Research,
University of Florida, 215 Williamson Hall, \mbox{P.O. Box 118440}, Gainesville, FL 
 32611-8440.}
\author{M. Chaparala}
\address{National High Magnetic Field Laboratory, Florida State University,
1800 E. Paul Dirac Drive, Tallahassee, FL  32306-3016.}
\author{Th. Jolic\oe ur}
\address{Service de Physique Th\'{e}orique, CEA Saclay, F-91191
Gif-sur-Yvette, France.}
\author{B.H. Ward and D.R. Talham}
\address{Department of Chemistry, University of Florida, 200 Leigh Hall,
P.O. Box 117200, Gainesville, FL  32611-7200.}
\date{\today}
\maketitle
\begin{abstract}
	The magnetic susceptibility of the $S = 2$ quasi-linear chain Heisenberg
antiferromagnet  \mbox{(2,$2'$-bipyridine)trichloromanganese(III)}, $MnCl_{3}(bipy)$, has been
measured from 1.8 to 300 K with the magnetic field, $H$, parallel ($\|$) and 
perpendicular ($\bot$) to the chains.  The analyzed data yield $g\approx 2$ 
and $J\approx 35$ K.  The magnetization, $M$, has been studied at
\mbox{30 mK} and
1.4 K in $H$ up to \mbox{16 T}.  No evidence of long-range order is 
observed.  Depending on crystal orientation, $M\approx 0$ at \mbox{30 mK} until
a critical field is achieved ($H_{c\|} = 1.2\pm 0.2{\text{ }}$T and 
$H_{c\bot} = 1.8\pm 0.2$ T), where $M$ increases 
continuously as $H$ is increased.  These results are interpreted as evidence
of a Haldane gap.
\end{abstract}
\narrowtext
	Ever since Haldane's prediction in 1983 \cite{1,2}, a significant amount
of experimental and theoretical effort has been focused on understanding the 
difference between half-integer and integer spin linear chain Heisenberg 
antiferromagnets.  These endeavors established the presence of the Haldane gap
$\Delta$ in \mbox{$S = 1$} systems, where \mbox{$\Delta_{S=1}/J=0.41$} and $J$ ($> 0$) is the 
nearest neighbor exchange energy.  Experimentally, 
$[Ni(C_{2}H_{8}N_{2})_{2}NO_{2}][ClO_{4}]$ \cite{3}, known as NENP, has received 
considerable attention since $J\approx45$ K is sufficiently large to access the 
Haldane phase with a variety of probes.  Along with inelastic neutron
scattering \cite{4,5,6} and high magnetic field ESR \cite{7,8} studies, 
magnetization investigations of single crystals of NENP at temperatures 
significantly below the gap have explicitly demonstrated the presence of 
$\Delta_{S=1}$  consistent with Haldane's description \cite{9,10,11}.  
Recently, several numerical studies have been performed on $S = 2$ 
antiferromagnetic quantum spin chains, and the results yield 
$\Delta_{S=2}/J = 0.08$\cite {12}, $0.055 \pm 0.015$ \cite{13}, $0.05$
\cite{14}, $0.085\pm 0.005$ \cite{15}, and $0.049\pm 0.018$ \cite{16}.   In 
addition, Schollw\"{o}ck and Jolic\oe ur \cite{15} have studied the phase 
diagram of the $S = 2$ spin chain Hamiltonian that includes single-ion ($D$) and
exchange ($\eta$) anisotropies. Such a Hamiltonian, with the addition of 
rhombic anisotropy, $E$-term, is
\begin{eqnarray}
{\cal H} =J\sum_{i} (S^{x}_{i}S^{x}_{i+1}+S^{y}_{i}S^{y}_{i+1}+
\eta S^{z}_{i}S^{z}_{i+1}) \nonumber \\
+D\sum_{i} (S^{z}_{i})^2 + E\sum_{i} [(S^{x}_{i})^2 -(S^{y}_{i})^2]. \label{H}
\end{eqnarray}
For $E = 0$, $\eta= 1$ and $D > 0$, the analysis of Schollw\"{o}ck and 
Jolic\oe ur indicates the Haldane phase 
is maintained only for $D/J \leq 0.04 \pm 0.02$.  For larger positive $D$ 
values, a gap-less phase with XY character separates the Haldane phase from a
gapped phase with a singlet ground state.  For large negative values of $D$, 
long range ferromagnetic or antiferromagnetic order is established, depending 
on the sign of $\eta$.  From their phase diagram, Schollw\"{o}ck and Jolicœ\oe ur 
\cite{15} conclude that it is unlikely to find real $S = 2$ materials that 
achieve the Haldane state. Nevertheless, attempts to find appropriate $S = 2$ 
systems have been reported, and as predicted, these materials appear to 
experience a transition to a long-range ordered state before reaching the 
Haldane phase \cite{17,18}.

	In this letter, we report the first evidence of the presence of the 
Haldane gap in an $S = 2$ chain antiferromagnet.  The magnetic properties of 
(2,$2'$-bipyridine)trichloromanganese(III), $MnCl_{3}(bipy)$, where $bipy$ represents
(2,$2'$-bipyridine), $C_{10}H_{8}N_{2}$, were first reported by Goodwin and 
Sylva \cite{19} down to \mbox {118 K} in 1967, and the crystal structure was published 
by Perlepes {\it et al.} \cite{20} in 1991.  The system consists of a quasi-linear 
chain of $Mn^{3+}$ ($S = 2$) ions in $MnCl_{2}(bipy)$ units, asymmetrically
connected by bridging $Cl^{-}$ ions (Fig.\ \ref{fig1}).  The $Mn^{3+}$
coordination is a distorted octahedron
where the $Mn-Cl$ bonds along the chain axis (\mbox{$Mn1-Cl1=2.51$ \AA } and 
$Mn1-Cl1'=2.71$ \AA) are longer than those in the equatorial plane (2.24 \AA )
\cite{20}.  Axial elongation results in singly occupied $dz^{2}$ orbitals and
the expectation that $\mid D \mid < \mid J \mid$ and
$D < 0$ \cite{21,22}.  The $Mn-Mn$ intrachain distance is 4.83 \AA , and the
strongest $Mn-Mn$ interactions are through overlap of the metal $dz^{2}$ orbitals 
with the chlorine bridge.  The $Mn-Cl-Mn$ bond angle of $135^{o}$ is consistent 
with antiferromagnetic exchange \cite{23}.  Furthermore, the closest $Mn-Mn$ 
distance between chains is 7.96 \AA , and since no ligands bridge
the chains, interchain exchange, $J'$, is expected to be small.  Following published
procedures \cite{20}, a total of four batches of the material have been prepared.
Two attempts provided powder-like material and the other two runs resulted
in samples of microcrystals with typical size of 
$90 {\text{ }}\mu m \times 0.4{\text{ }} mm \times 0.9{\text{ }} mm$.  The material was identified using 
combustion and single crystal X-ray analyses. 

The temperature dependence of the magnetic susceptibility,
$\chi (T)$, of approximately 2.4 mg of 90 oriented single crystals was studied 
from 1.8 to 300 K in a static
magnetic field of 0.1 T using a commercial SQUID magnetometer (Fig.\
\ref{fig2}).  
The single crystals were glued to weighing paper, 
using clear fingernail polish, before being placed in a gelcap which was held 
in a straw.  Although we could not measure the background of this system 
independently, a similar arrangement was constructed, sans sample, and measured.
This background contribution has been subtracted from the data reported in 
Fig.\ \ref{fig2}, where the data points at $300$ K are independent of orientation
and are normalized to the average value measured on powdered specimens,
weighing nominally 30 mg. The normalization was preformed to correct for a
small descrepancy between the measured and actual backgrounds.
Ultimately, our $\chi$($T$ = 300 K) value is consistent with the one reported previously \cite{19}. 

	The $\chi (T)$ results (Fig.\ \ref{fig2}) show a broad peak near 
\mbox{100 K}, anisotropy for $T<80$ K, a strong upturn at the lowest 
temperatures, but no indication of long range order. While the broad peak is 
the expected behavior for linear chain Heisenberg antiferromagnets, the 
Curie-like increase at low temperatures may be associated with impurities. The 
anisotropy is such that $\chi_{\bot} > \chi_{\|}$ which is in contrast to the 
$Cr^{2+}$ compounds\cite{17,18} where $\chi_{\bot} < \chi_{\|}$. Since no 
explicit expressions exist for $\chi (T)$ of an $S=2$ linear chain Heisenberg
antiferromagnet over a broad temperature range that includes anisotropy terms,
we have fit the data of Fig.\ \ref{fig2} to 	
\begin{equation}	
\chi (T)= \chi (0) + \frac{C}{T} +\chi_{\text{LCHA}}(S=2,g,J,T){\text{ ,}} \label{susc}
\end{equation}
where $\chi_{\text{LCHA}}(S=2,g,J,T)$ represents the $S = 2$ linear chain 
Heisenberg antiferromagnetic calculations of Weng \cite{25}, as parameterized 
by Hiller {\it et al.} \cite{26}. Since the expression for 
$\chi_{\text{LCHA}}(S=2,g,J,T)$ is not expected to be valid when significant
anisotropy is present or in a region where a gapped phase might exist 
({\it i.e.} $T<J$), the fitting procedure focused primarily on the region 
$T>80$ K, with
the exception that the Curie constant $C$ was adjusted to the low temperature
data. The results of the fit, shown by the solid line in Fig.\ \ref{fig2},
yield  $\chi (0) = 0.0 \pm 0.5$ memu/mol, $C = 47.5\pm 0.5$ memu K/mol,
$J= 34.8 \pm 1.6$ K and $g= 2.04\pm 0.04$. The Curie constant could be
explained by a small concentration of impurity spins. However, we want to be
careful about making this assignment and trying to subtract this ``Curie-tail''.
For example, as previously mentioned, we know that $\chi_{\text{LCHA}}(S=2,g,J,T)$
is an inadequate description of $\chi (T)$ in this region. Nevertheless it is
noteworthy that various attempts to subtract a reasonable Curie-like
contribution always give $\chi (T)\rightarrow  0$ as $T \rightarrow 0$. To
further explore the nature of the magnetic signal at the lowest temperatures,
standard 9 GHz ESR was performed on a 
packet of 5 oriented crystals from 4 to $60$ K.  The observed signal is consistent 
with a concentration of approximately $0.05\pm  0.03 \% $ $Mn^{2+}$ spins 
($S = 5/2$, $g = 2$) that follow a Curie temperature dependence. The signal may
also contain contributions from trace amounts (at the ppm level) of $S=3/2$ and
$S=1/2$ extrinsic impurities. However, since the concentration of
ESR visible spins is more than an order of magnitude smaller than needed to
explain the static susceptibility data, we consider isolated $Mn^{3+}$ ions 
not in the chain environment and $S=1$ end-chain spins\cite{13,15}
as the most likely source of the low temperature behavior.
 
	Using a single crystal $Si$ cantilever magnetometer \cite{27} 
mounted in the mixing chamber of a commercial dilution 
refrigerator, the magnetization of several different single crystals 
has been studied at \mbox{$30\pm 5$ mK} and $1.4\pm 0.2$ K in magnetic fields up to
16 T. Typical
results at 30 mK are shown in Fig.\ \ref{fig4}, where the data were taken on two single 
crystals with approximate mass of 200 $\mu$g. A weak temperature 
independent diamagnetic background and a small constant magnetization consistent
with the saturation of impurity spins have been subtracted \cite{28}.  The two 
crystals were held together on the cantilever using a trace amount of vacuum 
grease.  The data taken at \mbox{$1.4$ K} are consistent with the results reported in 
Fig.\ \ref{fig2} and magnetization data taken at 2 K with the SQUID
magnetometer to be detailed elsewhere.  The results at $30$ mK (Fig.\ \ref{fig4}) clearly indicate the 
presence of a non-magnetic state at the lowest fields, and the continuous onset 
of a magnetic state for fields above a critical field, $H_{c}$, which depends 
on crystal orientation, namely \mbox{$H_{c\|}= 1.2 \pm 0.2$ T} and 
\mbox{$H_{c\bot} = 1.8 \pm 0.2$ T}.  Furthermore, no discontinuities, hysteresis, or 
other unexplainable anomalies \cite{28} were observed in $M$ vs. $H$ traces up to 
16 T at either 30 mK (inset of Fig.\ \ref{fig4}) or \mbox{1.4 K}.

	These results indicate no evidence for three dimensional long-range
 magnetic order to 30 mK, where either hysteresis or a spin-flop 
transition should have been detectable.  In addition, the presence of a critical
field, marking a transition from a non-magnetic to a magnetic state, for both
crystal orientations is incompatible with the existence of long-range order or 
the presence of a gap-less phase\cite{15}.  Consequently, the spins are either in
a Haldane or large $D/J$ phase.  However as discussed earlier, the crystal 
structure indicates that the single-ion anisotropy should be small compared to 
$J$.  Ergo, we conclude the critical fields are related to a Haldane gap, where
anisotropy in $H_{c}$ is related to a small $D \neq 0$ (Eq.\ (\ref{H})).  For the
present discussion, we will assume any rhombic anisotropy ({\it i.e.} the last 
term in
\mbox{Eq.\ (\ref{H}))} to be negligible, and we follow the type of analysis that has been 
applied successfully to NENP \cite{11,29}.  In other words, for $D = 0$, the 
singlet ground state is separated from the first excited states, which form a 
degenerate triplet, by the Haldane gap $\Delta_{S=2}$  \mbox{(Fig.\ \ref{fig5})}.  
The presence of $D \neq 0$ causes the triplet to split, with the estimates of the resultant energy gaps 
given by $\Delta_{\perp} = \Delta_{S=2}-2D{\text{ and }}\Delta_{\parallel} =
\Delta_{S=2}+4D$ as shown in Fig.\ \ref{fig5} \cite{29}.  Finally,  
application of an external magnetic field causes the $|1,\pm 1>$ state to
split, where the $|1,-1>$ level crosses the ground
state at a critical magnetic field given by
$g_{\parallel}\mu_{B}H_{c\parallel}=\Delta_{\perp}{\text{ and }} 
g_{\perp}\mu_{B}H_{c\perp}=\sqrt{\Delta_{\perp}\Delta_{\parallel}}$\cite{29,33}.
  Taking $g_{\|}$ and $g_{\bot} = 2.04\pm 0.04$, we obtain 
\mbox{$D =  0.3 \pm 0.1$ K} and $\Delta_{S=2} = 2.3 \pm 0.8$ K.  
With these values and $J= 34.8\pm 1.6$ K, we may obtain 
\mbox{$D/J=0.010\pm 0.003$} and \mbox{$\Delta_{S=2}/J = 0.07 \pm 0.02$}, 
which agree well with the numerical work\cite{12,13,14,15,16}. It is
interesting to note the sign of $D$, obtained from the analysis of the
critical fields, conflicts with the sign predicted by the crystal structure. 
However, both the numerical work and this experimental estimate should include 
an $E$-term.  In addition, the quasi-linear ({\it i.e.} ``corrugated'') nature 
of the $Mn$-chains \mbox{(Fig.\ \ref{fig1})} may generate additional energy 
shifts and magnetic field dependences \cite{31,32}. Finally, we comment that 
these additional effects may give rise to the anisotropy in 
$\chi (T<80{\text{ K}})$, see Fig.\ \ref{fig2}, which remains unexplained.

	In summary, our magnetic studies of the $S = 2$ quasi-linear chain 
Heisenberg antiferromagnetic system $MnCl_{3}(bipy)$ indicate the absence of 
any long range order down to $30$ mK.  At the same time
no magnetization is observed at 30 mK until a critical magnetic field is 
achieved in either orientation \mbox{($H_{c\|} = 1.2 \pm 0.2$ T} and 
$H_{c\bot} = 1.8\pm 0.2$ T), after which a net magnetization evolves continuously 
as the field is increased.  These results are interpreted as evidence 
of a Haldane gap, $\Delta_{S=2}$, and indicate experimentally that 
$\Delta_{S=2}/J = 0.07 \pm 0.02$, which is in good agreement with the
numerical results.  Additional experiments are needed to clarify the gap 
values and to understand the detailed role of the anisotropy present in this 
system.

	We are grateful for early sample synthesis efforts of B. Dodson and 
assistance with the ESR measurements from L.-K. Chou.  We acknowledge 
enlightening conversations with E. Dagotto and F. Sharifi. 
This work was made possible, in part, by funding from the National Science Foundation
through an individual research grant, DMR-9200671 (MWM), and through support of the 
National High Magnetic Field Laboratory (NHMFL) at Florida State University 
(FSU) in Tallahassee. Finally we thank J. R. Childress for making the magnetometer
(NSF DMR-9422192) available to us.

\begin{figure}
\caption{Left: The crystal structure of $MnCl_{3}(bipy)$ is shown.  
The important bond distances and angles are given in the text. Note a slight
corrugation of the line connecting successive $Mn^{3+}$ ions. Right: Magnetic 
orbital scheme showing the $Mn-Cl-Mn$ intrachain overlap.}
\label{fig1}
\end{figure}

\begin{figure}
\caption{The magnetic susceptibility, $\chi (T)$, measured in a magnetic field of
0.1 T applied perpendicular (O) and parallel (+) to the chains.  The solid 
line indicates a fit of the data to Eq.\ (\protect\ref{susc}) as explained in the text.  The inset
shows $\Delta \chi (T) =\chi _{\bot} (T) - \chi_{\|} (T)$, where typical 
uncertainty limits are indicated at high and low temperature.}
\label{fig2}
\end{figure}

\begin{figure}
\caption{$M(H)$ at 30 mK for the field applied for perpendicular (O) and
parallel (+) to the chains of two single crystals with total mass $\approx$
200 $\mu$g, where both traces are offset vertically from zero for clarity.
The arrows identify $H_{c\bot}$ and $H_{c\|}$ with their respective 
uncertainties. 
The inset shows a more complete view of the $H\bot$ chain data.} 
\label{fig4}
\end{figure}

\begin{figure}
\caption{The energy levels for the singlet (S) ground state and the first 
excited triplet (T) state are shown schematically for $E = 0$ (Eq.\
(\protect\ref{H})):  (a) 
for $D = 0$ and (b) for $D > 0$.  In (c), the horizontal axis is the externally 
applied magnetic field. The field dependence of the states are shown 
schematically for $H \|$ chains (solid lines) and $H \bot $ chains (broken 
lines).  The crossing of the $|1,-1>$ triplet state with the singlet ground state 
defines $H_{c}$.}
\label{fig5}
\end{figure}
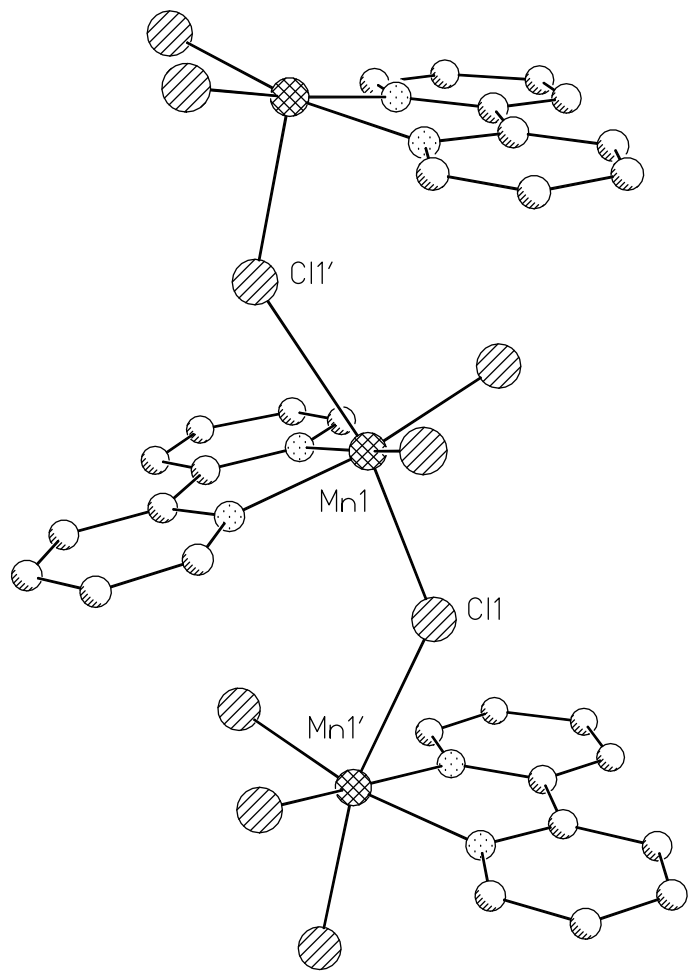
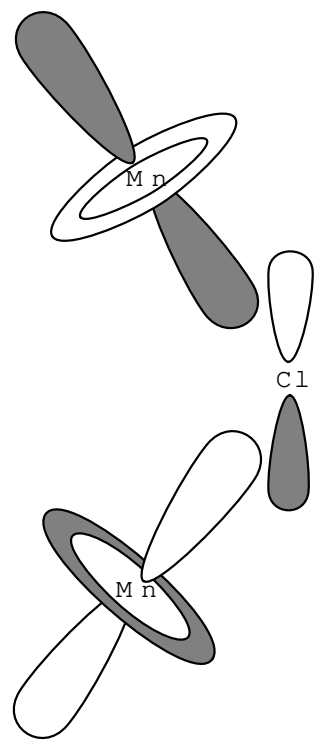
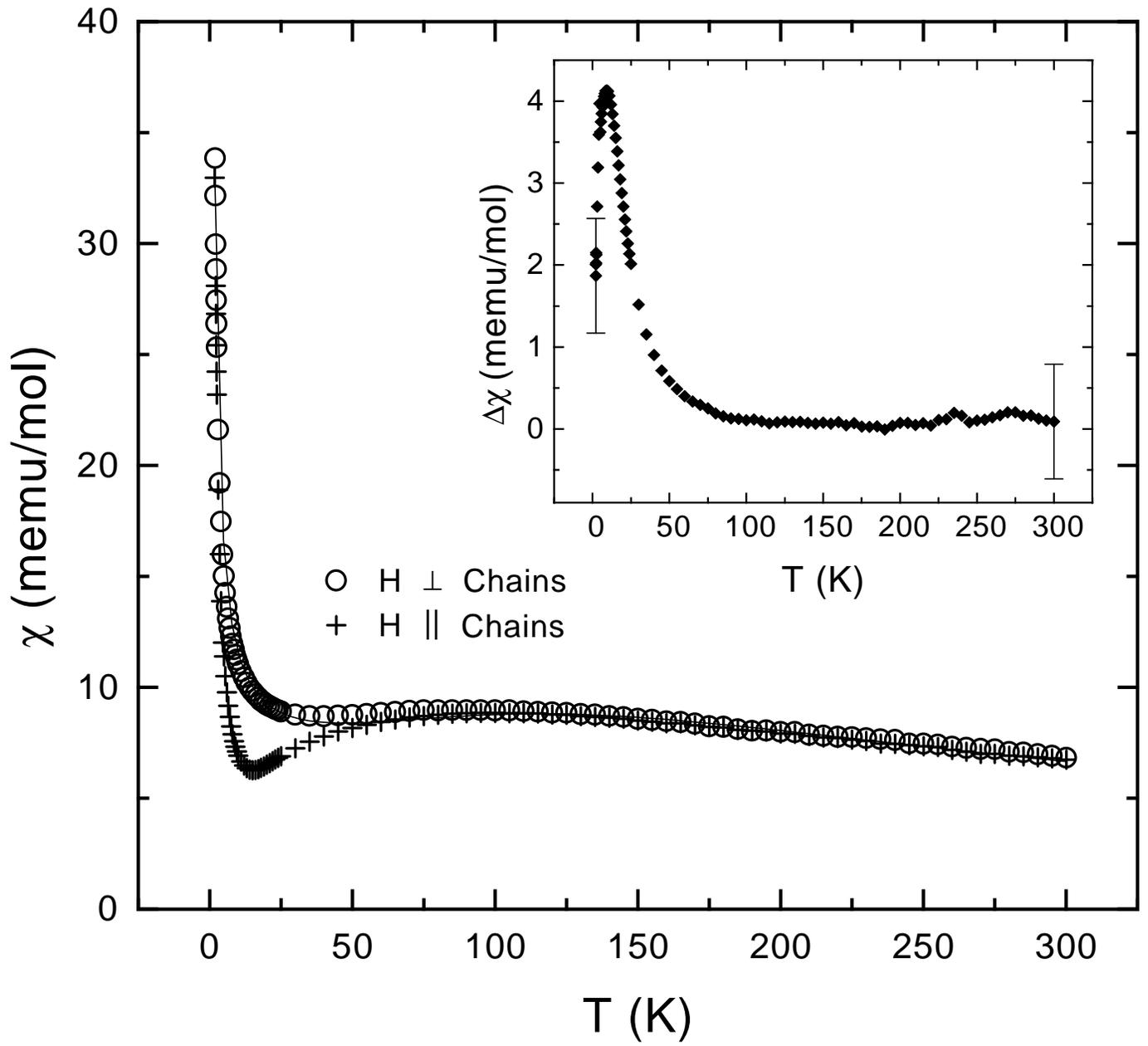
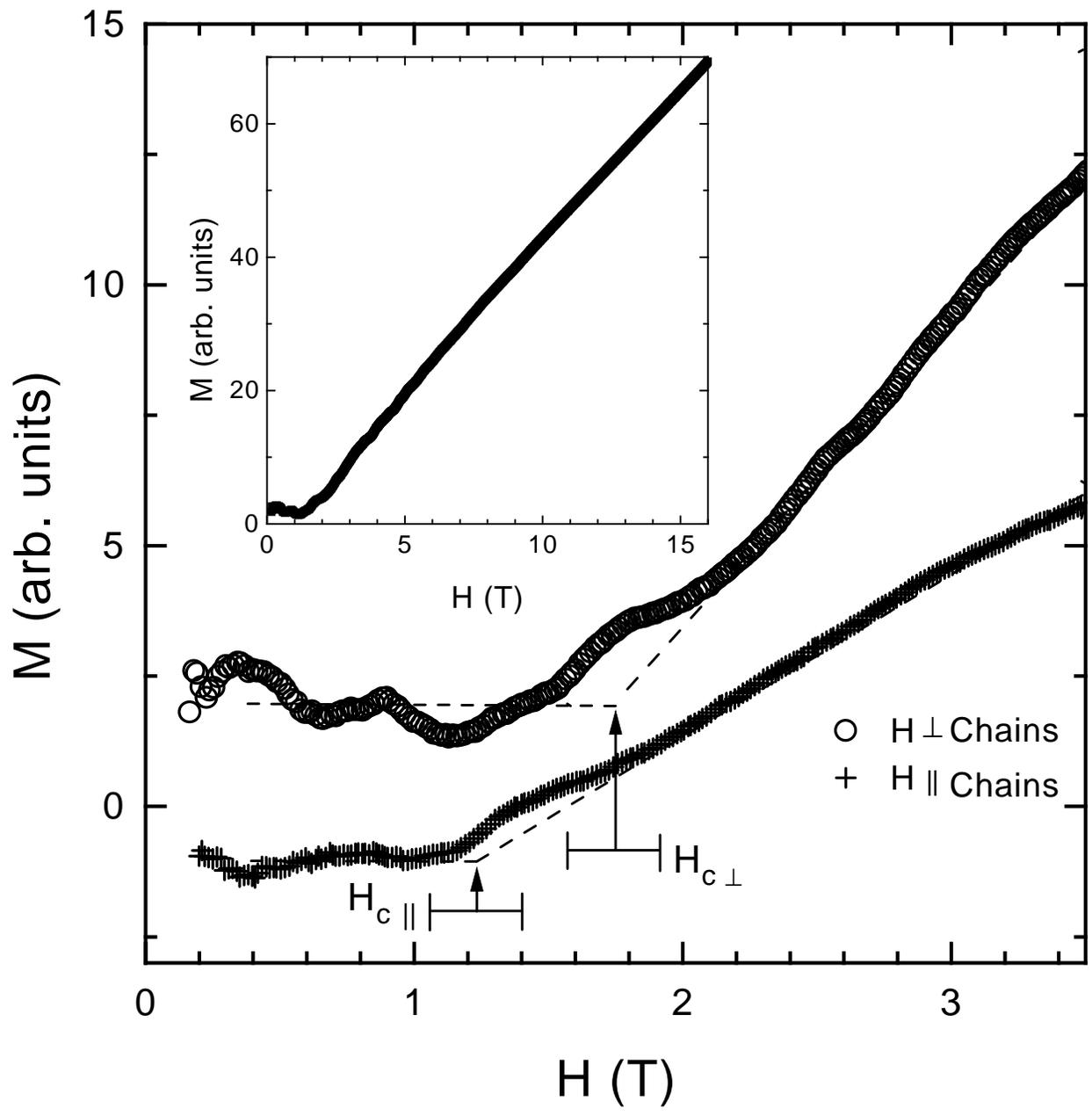
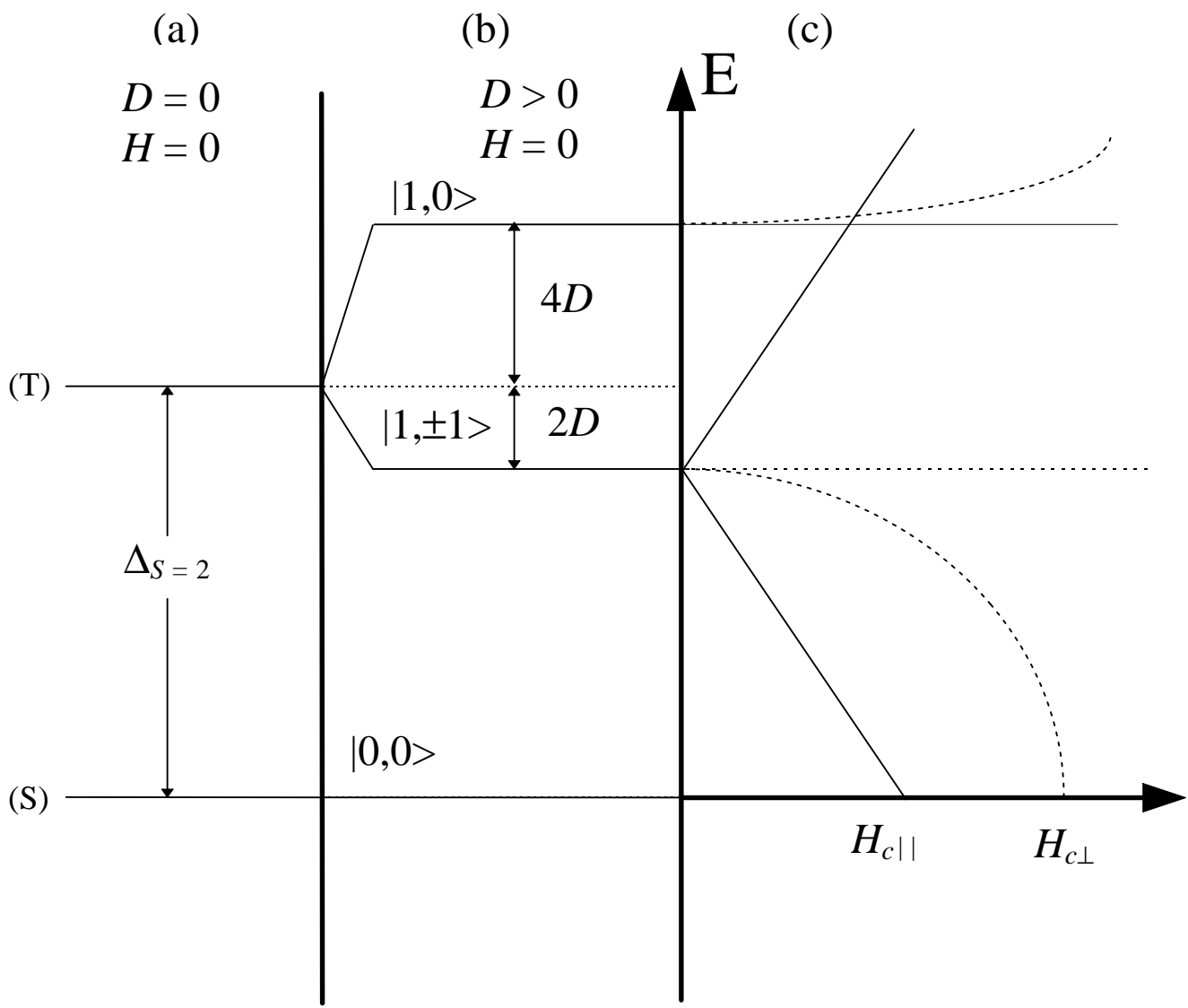
\end{document}